\documentclass[pre,twocolumn,amsmath,amssymb,aps, longbibliography]{revtex4-2}

\usepackage{graphicx}
\usepackage{dcolumn}
\usepackage{bm}

\begin{document}

\preprint{APS/123-QED}

\title{Self-assembly of two-dimensional, amorphous materials on a liquid substrate}

\author{Deborah Schwarcz}
\email{deborah.schwarcz@gmail.com}
\author{Stanislav Burov} 
 \email{stasbur@gmail.com}
\affiliation{Physics Department, Bar-Ilan University, Ramat Gan 5290002, Israel}

\date{\today}

\begin{abstract}
Recent experimental utilization of liquid substrate in the production of two-dimensional crystals, such as graphene, together with a general interest in amorphous materials, raises the following question: is it beneficial to use a  liquid substrate to optimize amorphous material production? 
Inspired by epitaxial growth, 
we use a two-dimensional coarse-grained model of interacting particles to show that introducing a motion for the substrate atoms improves the self-assembly process of particles that move on top of the substrate. We find that a specific amount of substrate liquidity (for a given sample temperature) is needed to achieve optimal self-assembly. 
Our results illustrate the opportunities that the combination of different degrees of freedom provides to the self-assembly processes.
\end{abstract}

\maketitle

\section{Introduction}
The rise of two-dimensional (2D) materials opens a variety of possibilities for materials science and nanotechnology \cite{amourphous_two_dimension,  2D_Nanophotonics,2D_aplication, Van_der_Waals_heterostructures, Novoselov10451, 2D_materials}.
  It is possible to distinguish between two different categories of 2D materials, crystal, and amorphous materials. While crystals have a periodic structure, amorphous materials are categorized by the lack of periodicity. Material microscopic structure
has a crucial impact on its global properties; therefore, controlling the self-organization of two-dimensional materials, such as ordered or disordered graphene, is vital to optimizing their performance \cite{Structure_of_graphene_and_its_disorders, different_feature_of_different_phase}.

Two-dimensional materials are frequently produced by bottom-up techniques like chemical vapor deposition (CVD), plasma-enhanced CVD , or physical vapor deposition  \cite{amourphous_two_dimension,
CVD_thecnics, CVD_graphen, CVD_book}. In these methods, atoms are deposited on a substrate, move on the substrate, interact, and self-assemble. The main challenge of creating 2D amorphous matter by bottom-up techniques is to obtain a large and defect-free cluster. Controlling growth parameters such as temperature, pressure, and substrate geometry enable to fit the outcome with high reproducibility~\cite{Zhang_17,Meng_16,Zhang_17,Chen_19,Meixner_23,Pyziak,Nurminen_1,Chen_19, Schwarcz_Burov_2019}.
 Recently, a liquid substrate was experimentally utilized for crystal growth. The lack of a crystallographic substrate has been observed to positively impact crystallization, i.e. larger crystal size. \cite{Geng_liquid_Cu,Zeng_20,Boeck_27,Zhang_28}. 
 It is also possible to promote  the rearrangement of atoms by utilizing methods like radiation \cite{PNAS_stone_waals_deffect,STONE_radiation,ultravolet_thermal_noise,Light-controlled_self-assembly,electron_beam},  electric/magnetic fields \cite{Directed_Self-Assembly} and heating \cite{Temperature_impact_in_Two_Dimensional_Self_Assembled_Network}. 
 The superposition of several of the mentioned methods and their effect on inter-atomic interaction also has been explored.  
 For example,  thermal activation and UV radiation  promote the rearrangement of atoms in glassy systems. \cite{ultravolet_thermal_noise}. Reviews of diverse experimental methods and simulation techniques for self-assembly of nanoparticles to large clusters can be found in \cite{Directed_Self-Assembly,2D_self_assembly_review}.

 The self-assembly is a generic name for a microscopic process that determines the spontaneous self-organization of the building blocks of the material. 
It can be naturally stimulated or modified by controlling experimental conditions \cite{Directed_Self-Assembly,  3D_roughnness}. 
Previously self-assembly was explored by croase grained models such as terrace ledge kink models  \cite{BCF_model, dGilmer_simulation_self_assembly},  Kardar-Parisi-Zhang (KPZ) equation \cite{KPZ},tile assembly models \cite{tam_introduction,BRUN_tam} and solid on solid (SOS) models \cite{Chatterjee2007,Pyziak,Biehl}. 

On the level of a single particle, self-assembly occurs due to inter-particle interactions. 
In general, particles present in the vicinity of local energy minima are separated by significant energy barriers. From time to time, particles experience abrupt transitions between these local minima. 
These transitions occur due to random fluctuations that enable the system to escape from a metastable state \cite{noise}. We generally address random fluctuations as noise. It is possible to distinguish between two kinds of noise effects  on the single particles, a uniform effect that, on average, affects all the particles similarly and a heterogeneous impact that will affect each particle differently. 
In this study, we develop a coarse-grained model of interacting particles (a generalization of the SOS model~\cite{Chatterjee2007,Pyziak,Biehl}) to explore the impact of different noises on the self-assembly processes of amorphous materials. 
We separately introduce two kinds of noises in the model. 
One type has the same impact on each atom in the system; we call it uniform noise. The other noise has a different effect on each atom in the system, and hence the name:  local noise. 
The temperature is assumed to be constant across the sample; therefore, it is a uniform noise. In contrast, atoms of the liquid substrate move differently through the sample; thus, their motion causes a different substrate arrangement at each point, i.e., a local noise. The emerging questions are: What is the impact of the various noises on self-assembly processes? 
Which noise is beneficial for 2D  self-assembly? What happens when a cohort of these noises is applied?  
This study explores these questions by simulating a self-assembly process of an amorphous cluster on top of a liquid substrate.  We use  Molecular Dynamics (MD) to describe the motion of substrate atoms and the Kinematic Monte Carlo (KMC) approach to address the self-assembly of particles on top of the substrate.  
 Voronoi tessellation representation of the substrate interlinks these two approaches.   
 In our model, the motion of substrate constituents perturbs the interaction of the assembling particles. 
 The effects of uniform and local noises are quantified by measuring the size and compactness of the obtained cluster.

 In this paper, we first study the impact of uniform and local noise on the self-assembly of 2D amorphous clusters. Our results suggest that increasing the noise (local or uniform) up to some level has a beneficial effect on self-assembly. 
 A specific non-zero noise  must be present in the system for the self-assembly to create a large and uniform cluster. 
 We show that while the energetic pathways (created by different noise types) are disparate on the single-particle level, the apparent result on the cluster formation is similar. 
 Both noises can amplify the self-assembled cluster.  

\section{MODELS AND METHODS}

We simulate two-dimensional, amorphous cluster growth on top of a liquid substrate. Our model consists of a substrate and particles that move and self-assemble on top of the substrate.  The substrate is  a set of atoms that can reorganize. 
Initially, the substrate atoms are randomly dispersed. To obtain more or less uniform substrate density, we divide the sample into equal squares and randomly introduce an atom into each square. The substrate atom number $i$ interacts with substrate atom number $j$ via Lennard-Jones potential $V_{i,j}$:
\begin{equation}
V_{i,j}=4\epsilon \left[ \left(\sigma /r_{i,j} \right)^{12}-\left(\sigma /r_{i,j} \right)^{6}\right] ,
\label{probability02}
\end{equation}
where $\epsilon=100$,  $\sigma=10^4$
and $r_{i,j}$ is the distance between atoms $i$ and $j$.  The surface is a square of size $20\sigma\times20\sigma$, periodic boundary conditions are implied. To consider only the short-ranged repulsion interaction between atoms, we cut off the potentials at $1.1\sigma$, i.e., WCA potential (see \cite{Chandler_WCA} ).
The overdamped Langevin equation determines the dynamics of the substrate atoms while Euler–Maruyama discretization method is utilized, i.e., the  position of the  $i$-ith particle is 
\begin{multline}
\vec{r_i} \left(t+\Delta t \right)=\vec{r_i} \left(t\right) +\sqrt{D\delta t} \left(\eta_x  \hat{x}+\eta_y\hat{y}\right) \\
   +\sum_{i=1}^{n}4\epsilon \left[ -12\frac{\sigma^{12}} {r_{i,j}^{13}} +6\frac{\sigma^6}{r_{i,j}^7}\right]\delta t   \hat{p}_{i,j}
\label{destination}
\end{multline}
where $D=1.5\times10^{-7}\sigma^2/\delta t$ is the diffusion coefficient, $n$ is the number of particles that are closer than $1.1\sigma$, $\delta t=0.005$ is the time step size and $\hat{p}_{i,j}$ is the unit vector in the direction $\vec{r_j}-\vec{r_i}$. $\eta_x$ and $\eta_y$ simulate Gaussian noise for each axis;
\small
\begin{equation}
\eta_{x/y}=\begin{cases}
    0.2355+2.0098\cdot b_{x/y},  & a_{x/y}<0.2645 \\
    2\cdot0.2355\cdot(b_{x/y}-0.5),&    0.2645\leq a_{x/y} \leq 0.7355 \\
    -0.2355-2\cdot0.0098 \cdot b_{x/y} ,&    a_{x/y} >0.7355
  \end{cases}
    \label{noises}
  \end{equation}
  \normalsize
where $a_{x/y}$ and $b_{x/y}$ are  uniformly distributed random numbers $\in$ $\left(0,1\right) $. The first four moments of $\eta_{x/y}$ coincide with corresponding moments of the Gaussian distribution.

\begin{figure}[t]
    \centering
\includegraphics[width=.8\linewidth]{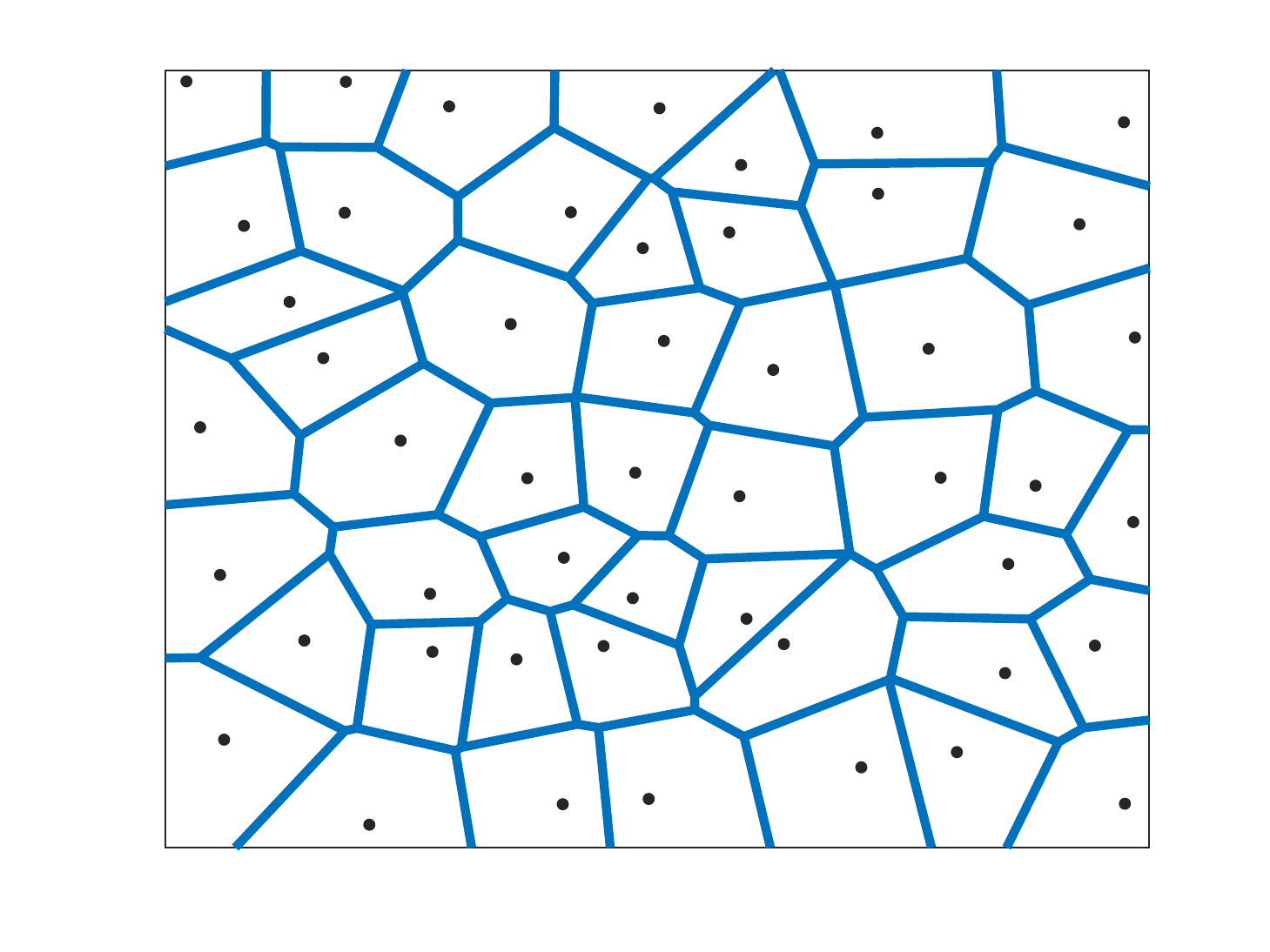}
  \caption {Vornoi tessellation description of the substrate; The points describe the substrate atoms that move via MD approach, i.e. Eq.\ref{destination}. The straight lines around each point determine the boundary of every Voronoi cell. 
  Each Voronoi cell can be occupied by up to one of the self-assembling particles that move on top of the substrate. Interaction energy of two self-assembling particles is determined by the length of the mutual boundary between the Voronoi cells that they occupy.
  } 
  \label{vornoi} 
\end{figure}

We use Voronoi tesselation \cite{Moukarzel_vornoi} to define the substrate sites created by the substrate atoms.  
 Voronoi tessellation is defined by a set of non-ordered sites, i.e., a set of randomly placed points. Each site defines a cell: Voronoi cell, that covers all the points that are closer to a given site than to any other site \cite{Moukarzel_vornoi}. For a given Voronoi cell, neighbor Voronoi cells are defined as cells that share a common boundary. The number of neighbor Voronoi cells and circumference length varies between different cells, see Fig.\ref{vornoi}.  
In our model, each  substrate atom represents the central point of a given Voronoi cell. Thus, the motion of substrate atoms alters the Voronoi tessellation.
 These modifications influence the geometry of all the cells simultaneously, but each Voronoi cell is affected uniquely. At the bottom line, these unique and random rearrangements introduce local noise to the system.    
During our simulation, we update the Voronoi tessellation. 
The number of MD steps between sequential Voronoi tessellation updates should be large enough for the change in the  structure to be significant.  That is, non-zero modifications for the cell circumference should be observed. But at the same time, many MD steps completely modify the Voronoi tessellation and disconnect previously neighboring cells.
To balance these criteria, we use a temporal step of $1800\delta t$ between sequential updates of the Voronoi tessellation.

So far, we have described the motion of substrate atoms. The substrate structure, i.e., Voronoi tessellation, defines the possible locations and dynamics of self-assembling particles that move on top of the substrate. Each  Voronoi cell can be occupied by up to one self-assembling particle. Initially, all the particles are randomly dispersed among the Voronoi cells.  
We use the KMC model to determine the transitions between different substrate sites of the self-assembling particles. 
At each iteration of the KMC, one particle can hop from one Voronoi cell to one of the unoccupied neighboring Voronoi cells.  
The local geometry of a given Voronoi cell determines the interaction energy of two nearby particles, i.e., self-assembling particles located at neighbor Voronoi cells. This 
  bonding interaction depends on the length of the mutual edge of the Voronoi cell; thus the total energy of a particle situated in Voronoi cell $s$ is provided by  
\begin{equation}
E_{s}=-J\frac{b_{s}\sum_{k}b_{k} f_{s,k}}{\sum_{k}f_{s,k}},
\label{adatom_energy}
\end{equation}
the summation is over all the neighbours of cell $s$ and $f_{s,k}$ is the length of the boundary between cells $s$ and $k$. $b_k$ is $1$ if cell $k$ is occupied and $0$ otherwise. 
This definition assumes that particle-particle interactions are linear with the distance between particles since the average distance between two randomly allocated particles in two adjacent Voronoi cells is proportional to the length of their mutual boundary \cite{Schwarcz_Burov_2019}.

The probability of a particle to attempt to leave its current site and jump to one of the adjacent empty cells follows
\begin{equation}
p_{s} =e^{\beta E_s}.
\label{Arrhenius}
\end{equation}
where  $\beta=\frac{1}{k_b T}$, $T$ is the temperature and $k_B$ is the Boltzmann constant.
If the attempt of the particle to leave its current cell $s$ is successful, it will consider all the potential destinations (i.e., empty neighboring cells). 
For each of those potential destinations the transition probability $p_{s\to k}$ is 
\begin{equation}
p_{s\to k}=\frac{e^{-\beta \Delta E_{s,k}}}{\sum_{k}e^{-\beta \Delta E_{s,k}}},
\label{probability03}
\end{equation}
where $\Delta E_{s,k}=E_{k}-E_{s}$ and the summation is over all the potential cells $k$. 
During the simulation, self-assembling particles locations are updated sequentially one after the other. 
Since the self-assembling particles are indistinguishable, it is possible to use sequential updating instead of random updating generally used in MC simulations \cite{sequential_updating}. 
Each $N$ KMC steps, the Voronoi tesselation is updated according to the algorithm described above.
In the following, we use the term vibration frequency $=1/N$ to describe the periodic updates of the Voronoi tessellation. 
These Voronoi tesselation updates are terminated after a specific (and vast) number of KMC steps. 
We allow the system to relax on top of  a specific (but randomly chosen) Voronoi tessellation.  
Notice that when such relaxation is introduced, we assume that we can control the motion of the substrate atoms. 
Such control is mathematically achieved via setting $D\to 0$ or rapid freezing of the substrate. 
Suppose the origin of the noise that affects substrate atoms rearrangements is achieved via an external source, such as rattling of the system. In that case, the relaxation phase occurs when this external source is switched off.   
We expand more about this relaxation phase in the next section.

The implementation of Voronoi tessellation as a substrate enables us to introduce the variations in the substrate during the self-assembly process. It allows studying the impact of substrate liquidity on the self-assembly process. 
It is worth noting that recently, a model describing ion transport processes used similar ideas of dividing the system into two coupled subsystems. MD represents one subsystem and the other by MC, see \cite{Coupled_MD_to_KMC}.

The main parameter that describes  "successful" self-assembly is cluster compactness.  To measure the cluster compactness, we sum over the lengths of the edges of all the cells.  The measure is defined by;
\begin{equation}
U=\frac{1}{2}\sum_{s}\sum_{k}f_{s,k}\left|(b_{s}-b_{k}\right)|,
\label{potential_energy}
\end{equation}
where $s$ indicates different particles and $k$ indicates the various neighbours of particle $s$.
Due to the presence of $\left|b_{s}-b_{k}\right|$ in Eq.~\ref{potential_energy},  the order parameter $U$ determined by the mutual edges of adjacent Voronoi cells, for situations where only one of the cells is occupied by a particle. 
Since the number of particles is fixed, small $U$ describes situations when many particles are clumped together and form clusters.  Large $U$ may be caused by holes in the cluster or the emergence of many small clusters instead of a large one. Situations when $U$ is large, are not desirable.

  \begin{figure}[t]
    \centering
\includegraphics[width=.80\linewidth]{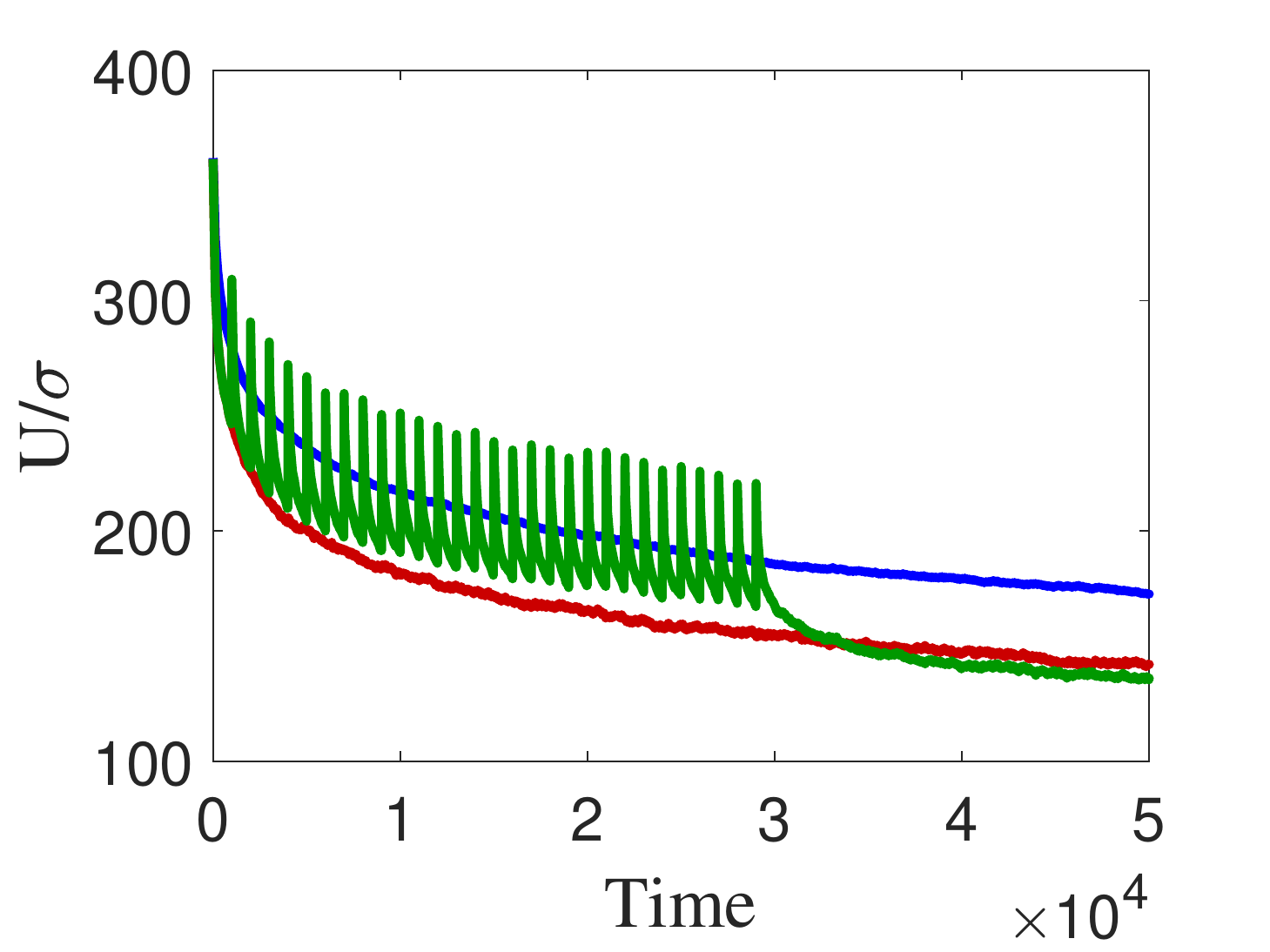}
  \caption {Cluster compactness of 260 particles as a function of time. Blue curve $k_B T/J=1/11$ and no substrate variations. Red curve $k_B T/J=1/7$ and no substrate variations. Green curve  $k_B T=1/7$ and dynamic substrate variations, vibration frequency is $10^{-4}$. The data was averaged over 100 realizations.}
  \label{threlineComp} 
\end{figure}

 \begin{figure}[t]
\begin{center}$
\begin{array}{ll}
\includegraphics[width=.42\linewidth]{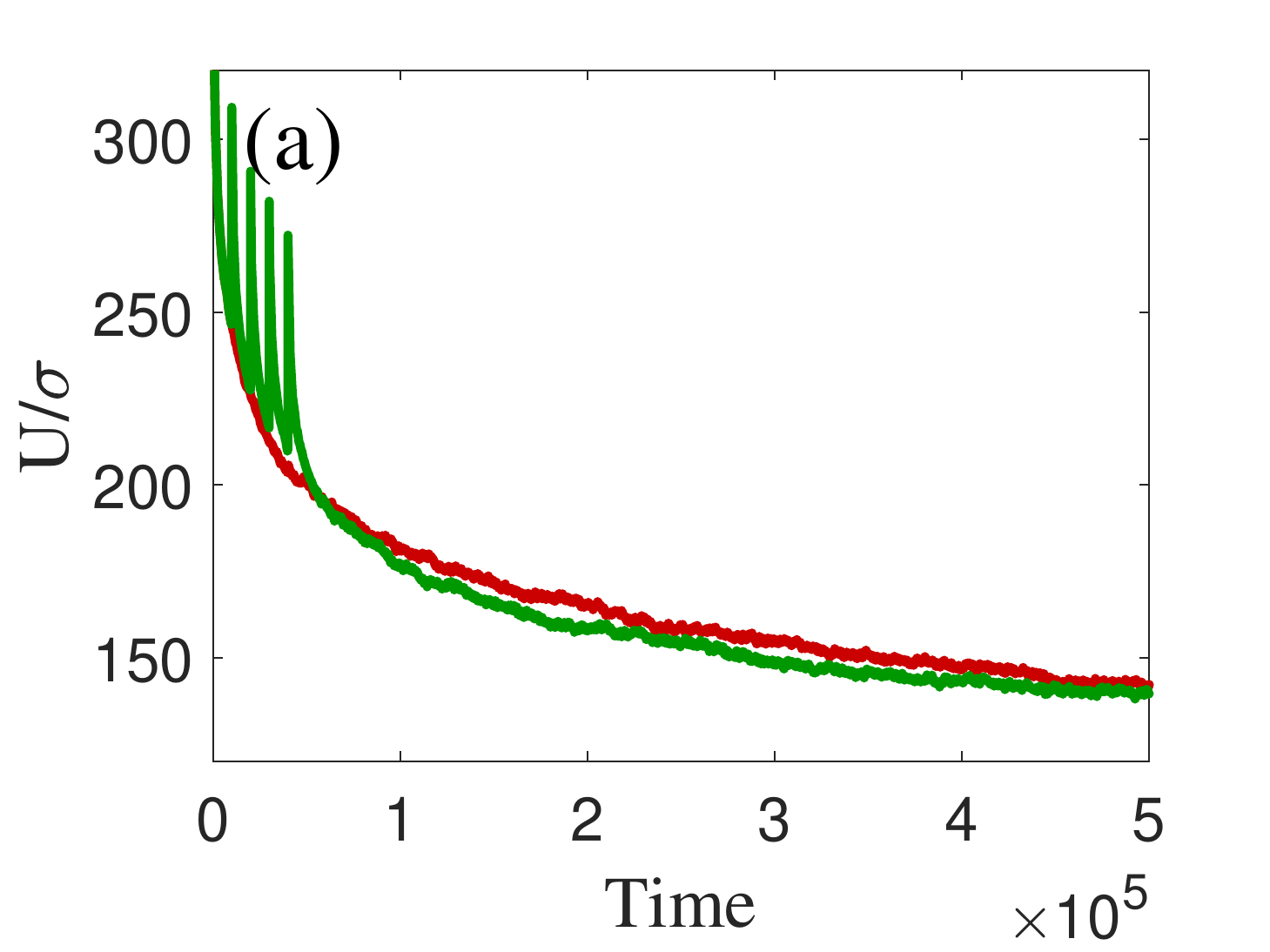}&
\includegraphics[width=.42\linewidth]{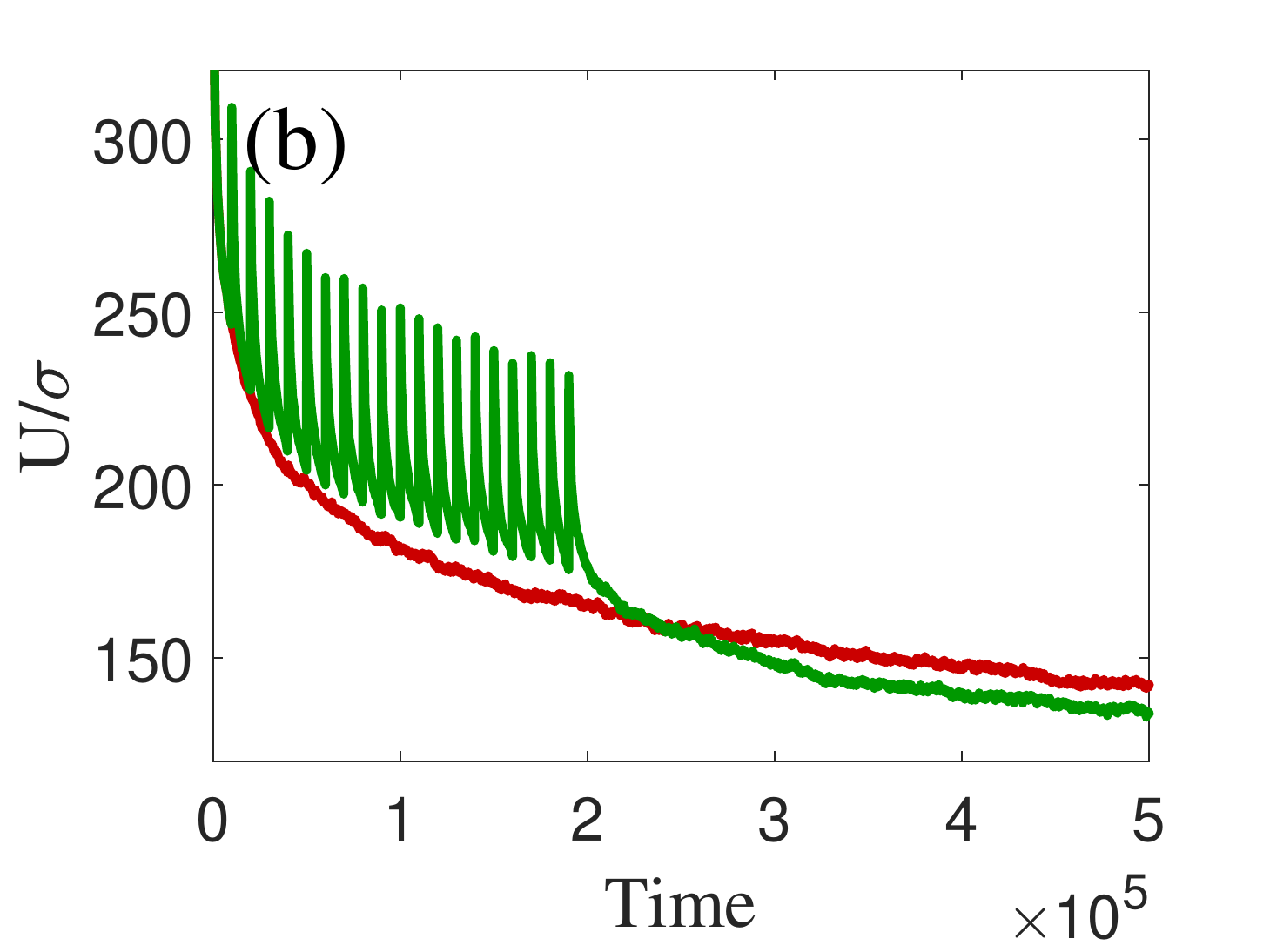}
\end{array}$
\end{center}

\begin{center}$
\begin{array}{ll}
\includegraphics[width=.4\linewidth]{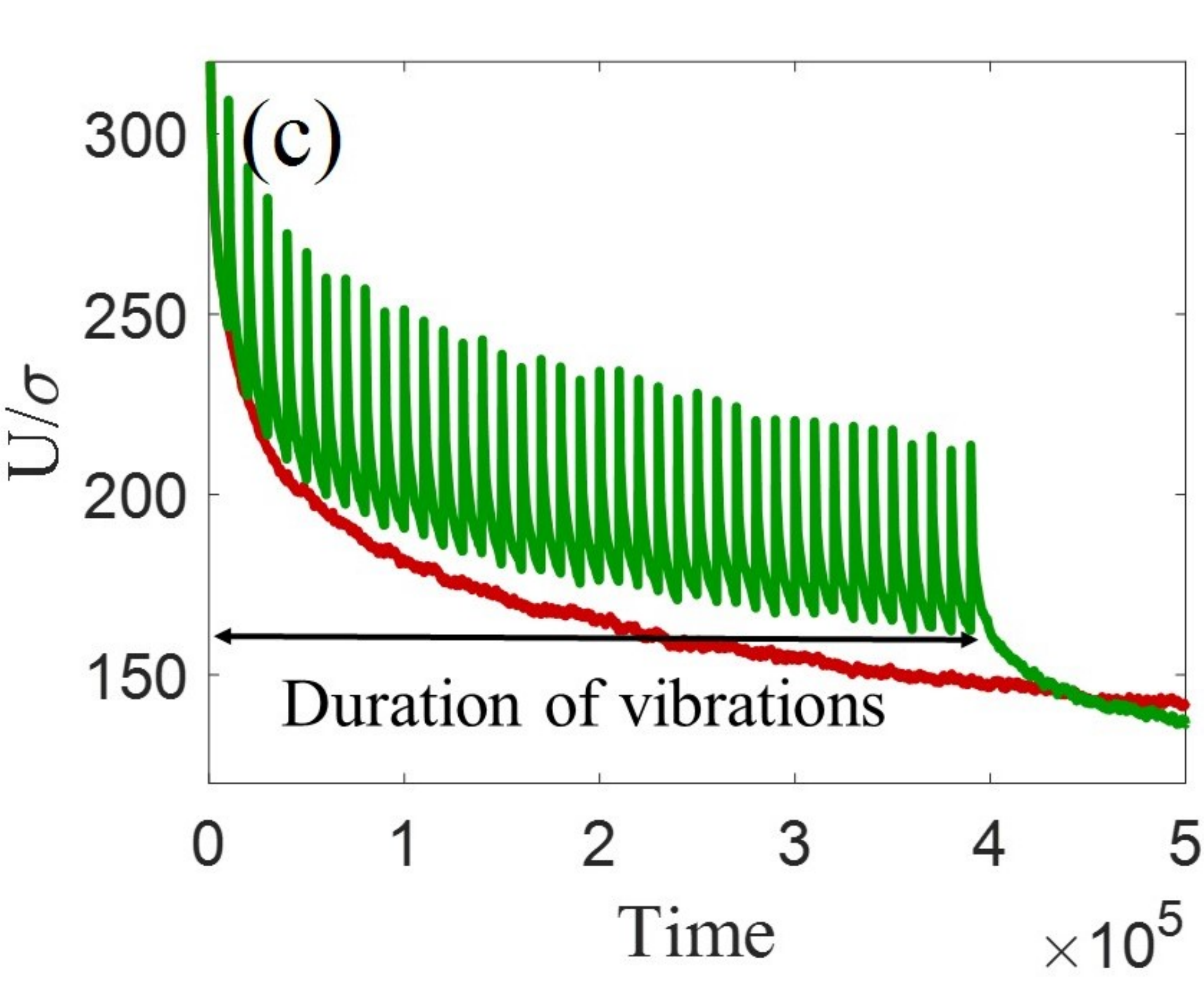}&
\includegraphics[width=.42\linewidth]{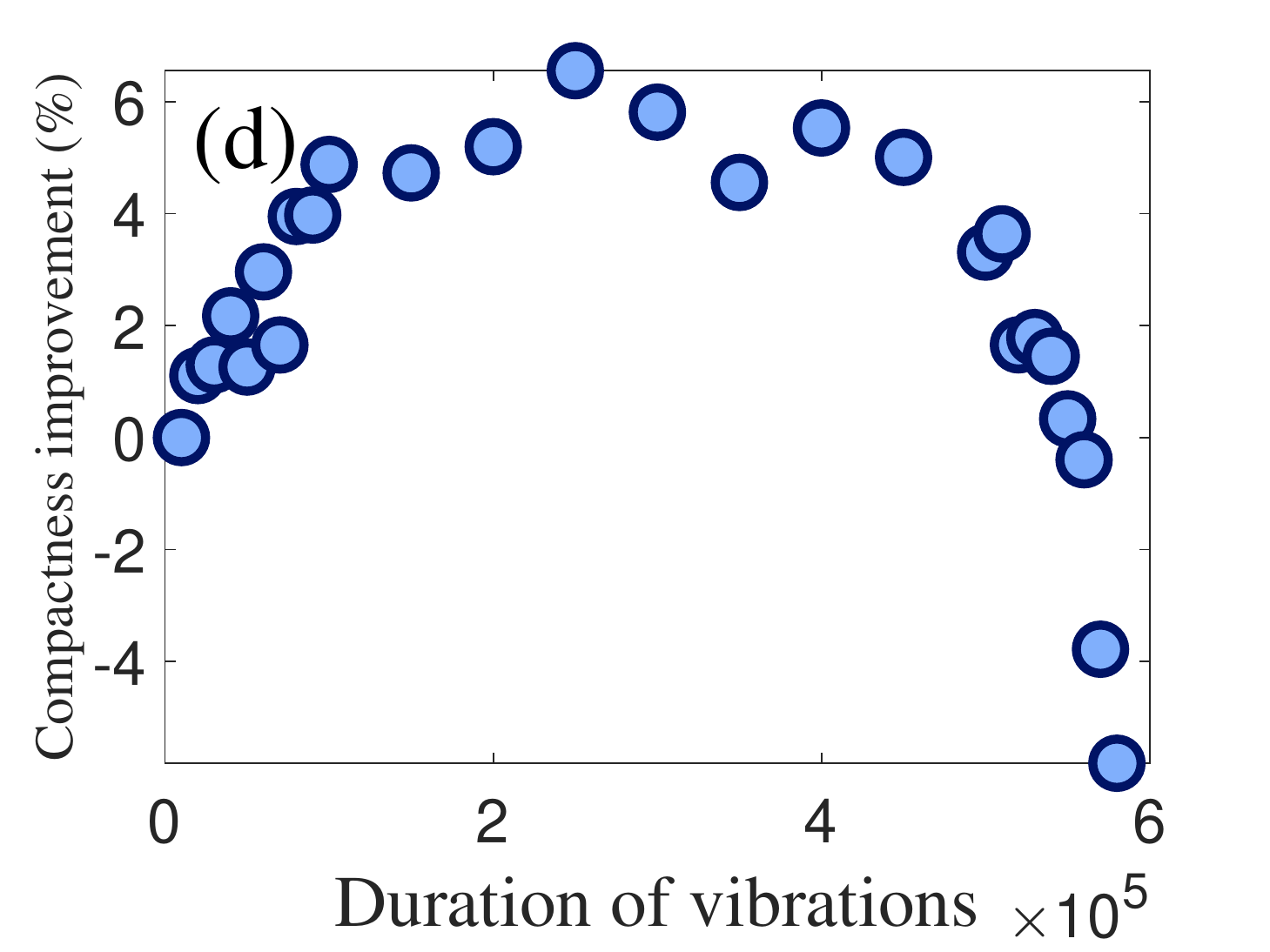}
\end{array}$
\end{center}
\caption{Panels {\bf{(a)}}-{\bf{(c)}} present the compactness us a function of time for $k_B T/J=1/7$, vibration frequency $10^{-4}$ and different periods of vibrations (illustrated in panel {\bf{(c)}}). Red curve describe the cases without substrate variations and green curves the cases where dynamical substrate variations were implemented. 
\textbf {a} Duration of vibration is $5\times 10^4$ KMC steps \textbf {b} Duration of vibration is $2\times 10^5$ KMC steps.  \textbf{c} Duration of vibration is $4\times 10^5$ KMC steps.
Panel \textbf{d} presents the improvement of the cluster compactness ($U$) for various duration of vibrations when the measurement time is $6\times 10^{5}$ and $k_B T/J=1/7$. 
The simulations were performed for $260$ particles and $100$ realizations.
}
\label{appendix_fig}
\end{figure} 

\section{Results}

Two types of noises are present in the model. The first noise is the thermal one that we term uniform. It is varied by controlling the temperature in Eqs.~(\ref{Arrhenius},\ref{probability03}). When the substrate is frozen, i.e., the periods between modifications of the Voronoi tessellation $\to\infty$, the behavior of $U$  is different for every $T$. 
In Fig.~\ref{threlineComp} $U$ of the self-assembling particles decays as a function of time. 
For two different temperatures, the minimal $U$s (achieved at the end of the measurement time) differ by $60\%$. This observation shows that self-assembly on a solid substrate is highly affected by temperature.
The second noise is the one that is introduced via rearrangements of the substrate atoms. The Voronoi cells differ pretty extensively from one place to another. Therefore the term "local noise" describes substrate variations. We vary the vibration frequency of the substrate. 
In Fig.~\ref{threlineComp} the $U$ with pronounced oscillations describes the situation when substrate rearrangements are introduced. The whole behavior looks as if the system is periodically rattled. 
As mentioned above, we stop these substrate variations after a specific amount of time (i.e., KMC steps). As shown in Fig.~\ref{threlineComp} after the substrate vibrations are terminated, $U$ starts to decay (on average).
Surprisingly, this decay leads to terminal values of $U$ that are smaller than the values achieved without substrate deformations, given that the measurement time is the same for both cases. 
This effect occurs even though at the start of the final relaxation, $U$ for the case with local noise is larger (i.e., less compact cluster) than $U$ for the case without such noise. 
It becomes clear that some balance should exist between the duration of substrate modifications, i.e., duration of vibrations and the period of the relaxation phase. 
In Fig.~\ref{appendix_fig} we present three representative cases of how $U$ behaves for different duration of vibrations (panels {\bf{(a)-(c)}}). 
Panel {\bf{(d)}} summarizes the findings for all possible values of duration of vibrations, from $0$ to the  measurement time. 
When the vibration duration reaches $\sim 1/3$ of the measurement time, the improvement in the cluster compactness ($U$) hits a plateau. This improvement disappears when the vibration duration is close enough to the measurement time and the period of the relaxation phase is too short. 
In the extreme cases when the duration of vibrations $\to$ measurement time, the improvement is negative, meaning that the obtained $U$ is larger than $U$ for the situation where no substrate variations are present. 
Enlargement of $U$ for a short period of relaxation agrees with the naive assumption that adding more noise to the system damages the chances of creating a compact cluster.  
Since we search for improvement of cluster formation, we use $\sim 1/3$ of the measurement time for the duration of vibrations.

\begin{figure}[t]
\begin{center}$
  \begin{array}{ll}
\includegraphics[width=.4\linewidth]{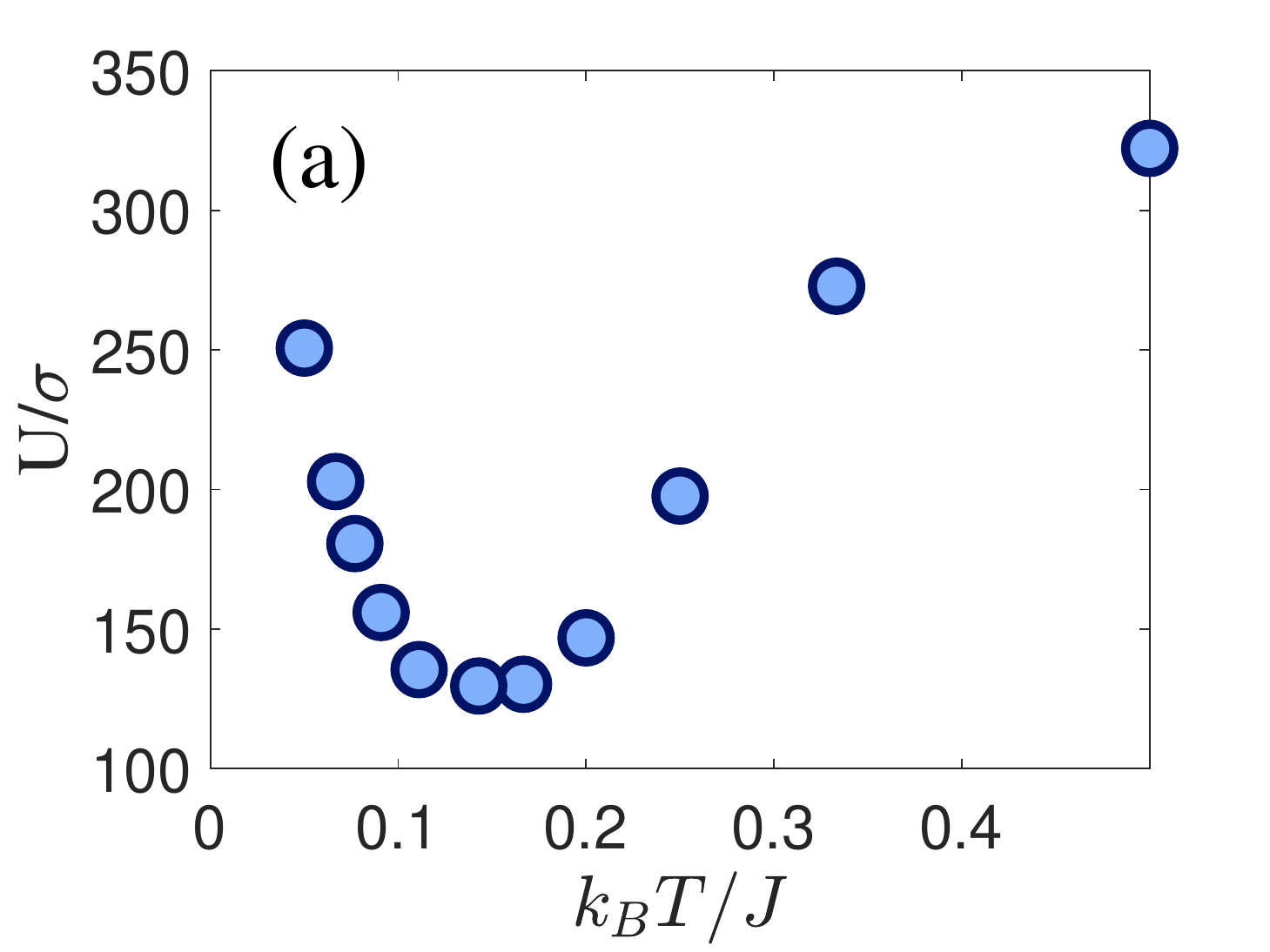}&
\includegraphics[width=.4\linewidth]{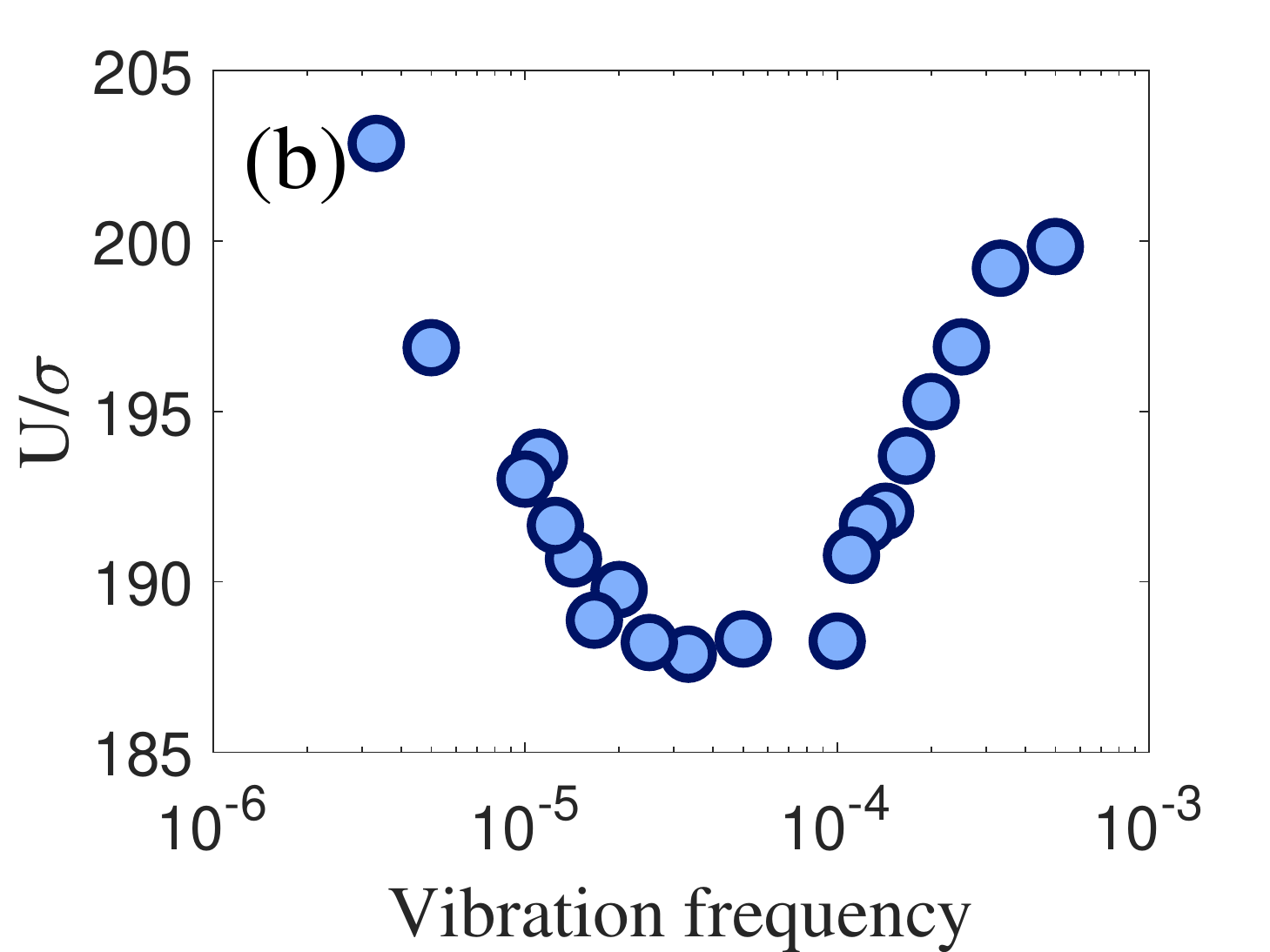}
  \end{array}$
  \end{center}
  
  \begin{center}$
\begin{array}{lll}
\includegraphics[width=.32\linewidth]{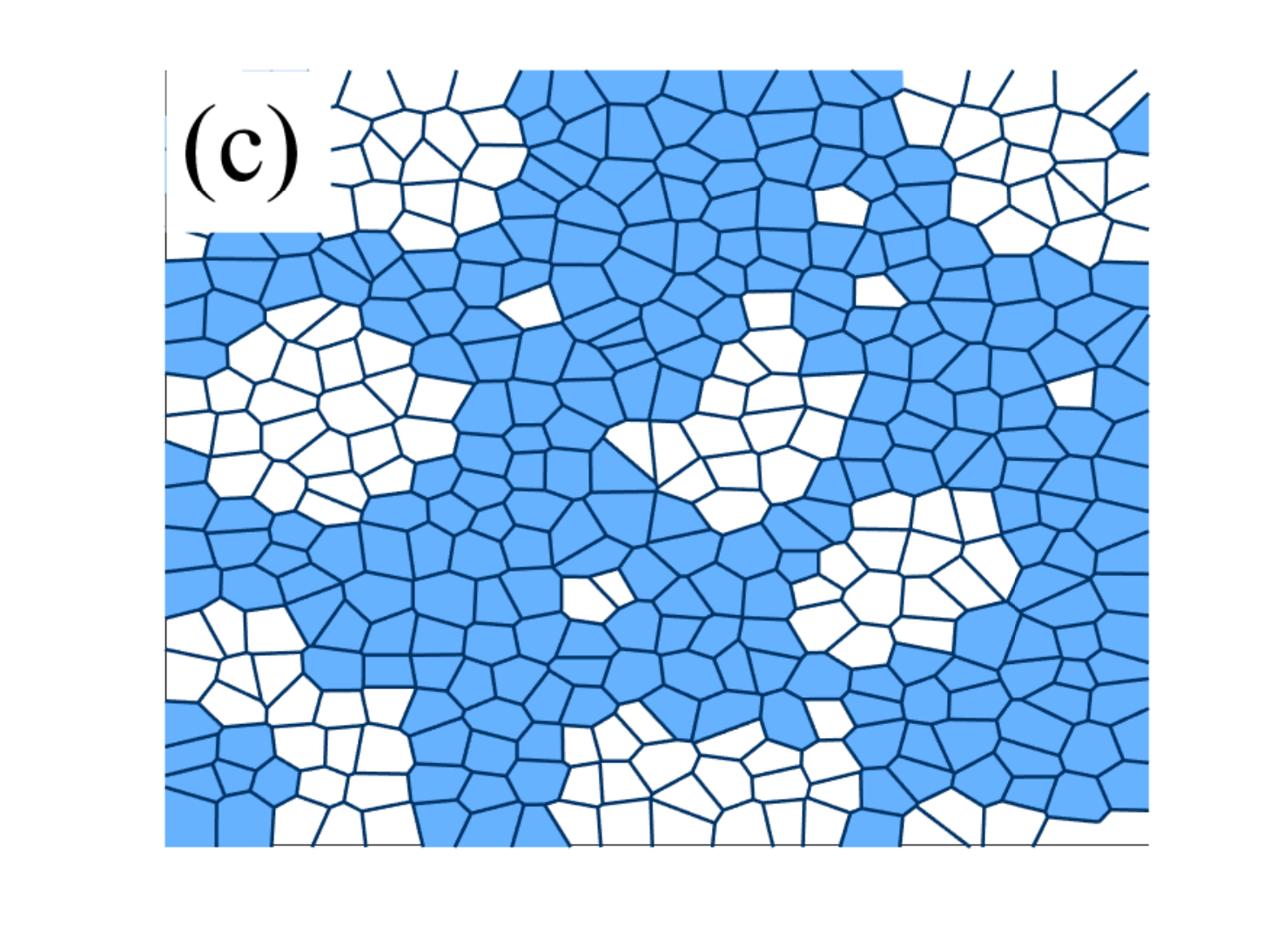}&
\includegraphics[width=.32\linewidth]{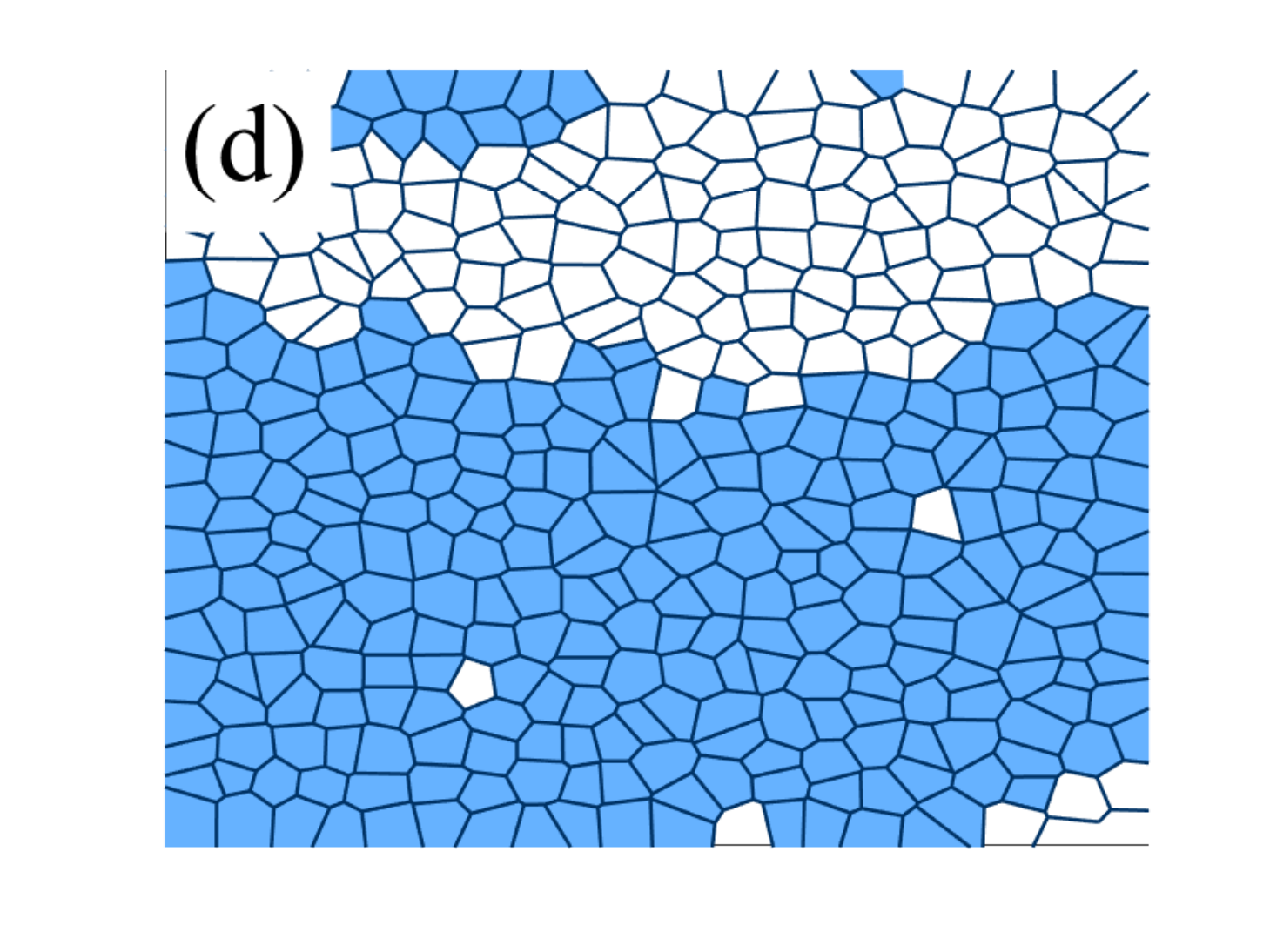}
\includegraphics[width=.32\linewidth]{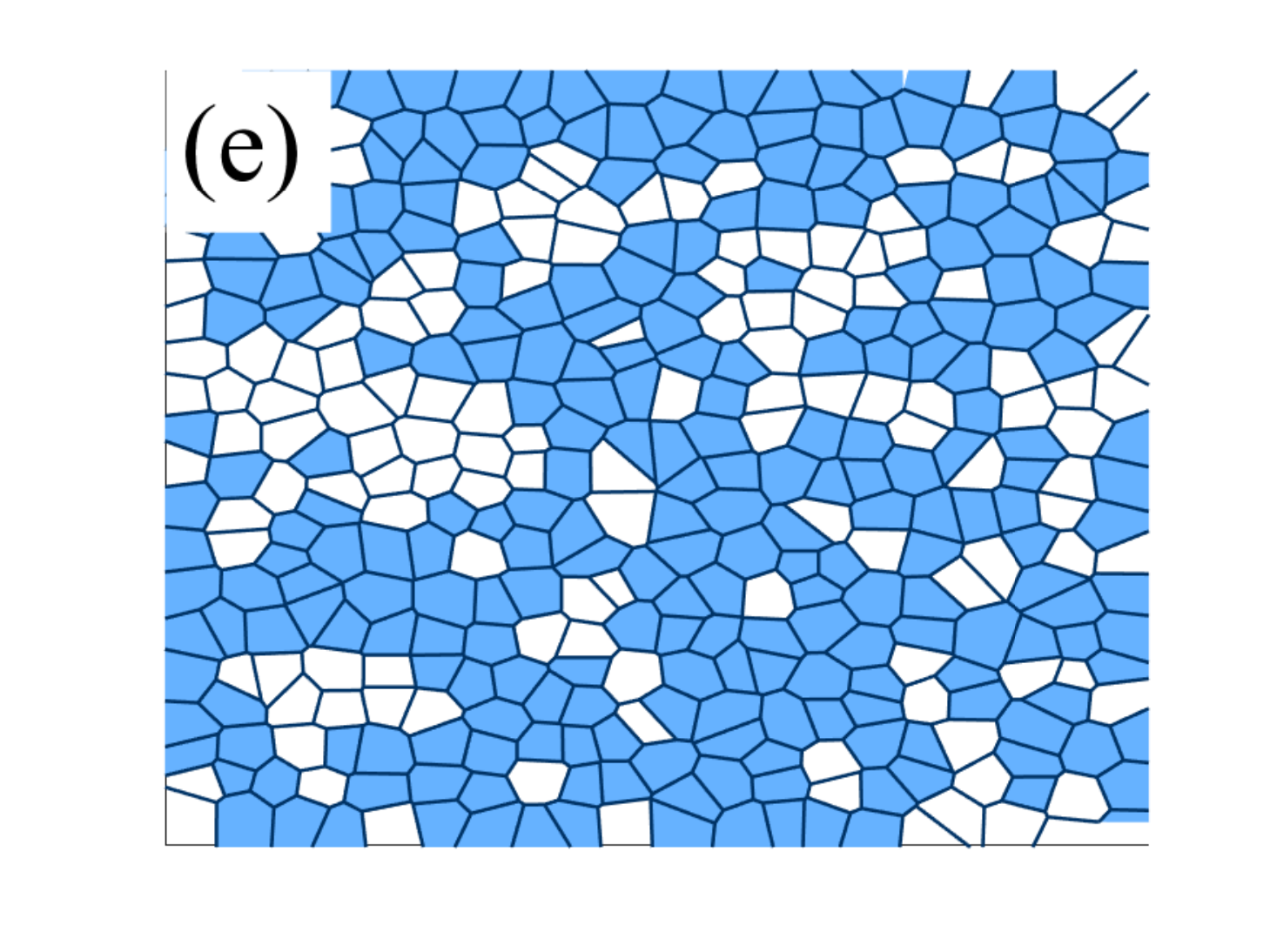}
\end{array}$
\end{center}

  \caption{
  Panels {\bf{(a)}}-{\bf{(b)}} display the cluster compactness as a function of the amount of "noise" in the system, for measurement time $10^6$ KMC steps .
  \textbf{a} Only thermal fluctuations are present. 
  \textbf{b} Thermal fluctuations and substrate variations are present the temperature is fixed, $k_B T/J = 1/15$.
  Panels {\bf{(c)}}-{\bf{(d)}} display snapshots of the formed cluster for measurement time $6\times 10^6$ KMC steps and and no substrate variations. \textbf{c} $k_B T/J = 1/15$. 
  \textbf{d} $k_B T/J = 1/7$. 
  \textbf{e} $k_B T/J = 1/3$. 
  Averaging over $99$ realizations was performed, while $260$ were used in each realization.
 }
  \label{U_vs_temp_or_vibre} 
\end{figure}

The two noises affect quite differently the immediate evolution of the system. 
While particles temperature increase leads to consistent small fluctuations, each substrate vibration violently destabilizes the system due to the reconfiguration of inter-substrate energetic bonds.
When we eliminate substrate variations and consider only  the effect of temperature, i.e., uniform noise, two distinct regimes appear.   For low temperatures, the system is stuck in a metastable state (Fig.~ \ref{U_vs_temp_or_vibre}  {\bf{(c)}} where  large "holes" persist for extremely long times in the cluster.
On the other hand, for high temperatures, the system stays in a homogeneous phase. Local  formation of cluster grains are  disassembled very fast (Fig.~\ref{U_vs_temp_or_vibre} {\bf{(e)}}). In both high and low-temperature limits, $U$ is large.
There is an optimal intermediate temperature where some balance is reached between the tendency to break loose and the opportunity to stay locally connected.
For this optimal temperature $U$ reaches a minimal value, as is shown in Fig.~\ref{U_vs_temp_or_vibre} {\bf{(e)}} and a compact cluster is obtained (Fig.~\ref{U_vs_temp_or_vibre} {\bf{(d)}}).
This result agrees with previous findings where an optimal interaction that leads to an efficient self-assembly process was observed~\cite{Bisker_pnas, Interactions_Optimized_Salvatore}.

When the temperature is set to be constant, and the vibration frequency is modified, 
a quite similar effect is observed. 
Fig.~\ref{U_vs_temp_or_vibre} {\bf{(b)}} shows that $U$ behaves non-monotonically with vibration frequency. 
There is a distinctive minimum of $U$ for the intermediate value of the vibration frequency. 
The small vibration frequency of the substrates acts beneficially, up to a specific limit. Further inclusion of additional noise is destructive for cluster formation.
Comparison of panels {\bf{(a)}} and {\bf{(b)}} in Fig.~\ref{U_vs_temp_or_vibre} shows that the roles of vibration frequency and temperature are close. 
The dynamic range of the $y$-axis in Fig.\ref{U_vs_temp_or_vibre} {\bf{(a,b)}} discloses that the size of the impact of the noises is different: temperature has much more effect on cluster compactness than substrate variations. 

\begin{figure}[t]
\begin{center}$
\begin{array}{ll}
\includegraphics[width=.4\linewidth]{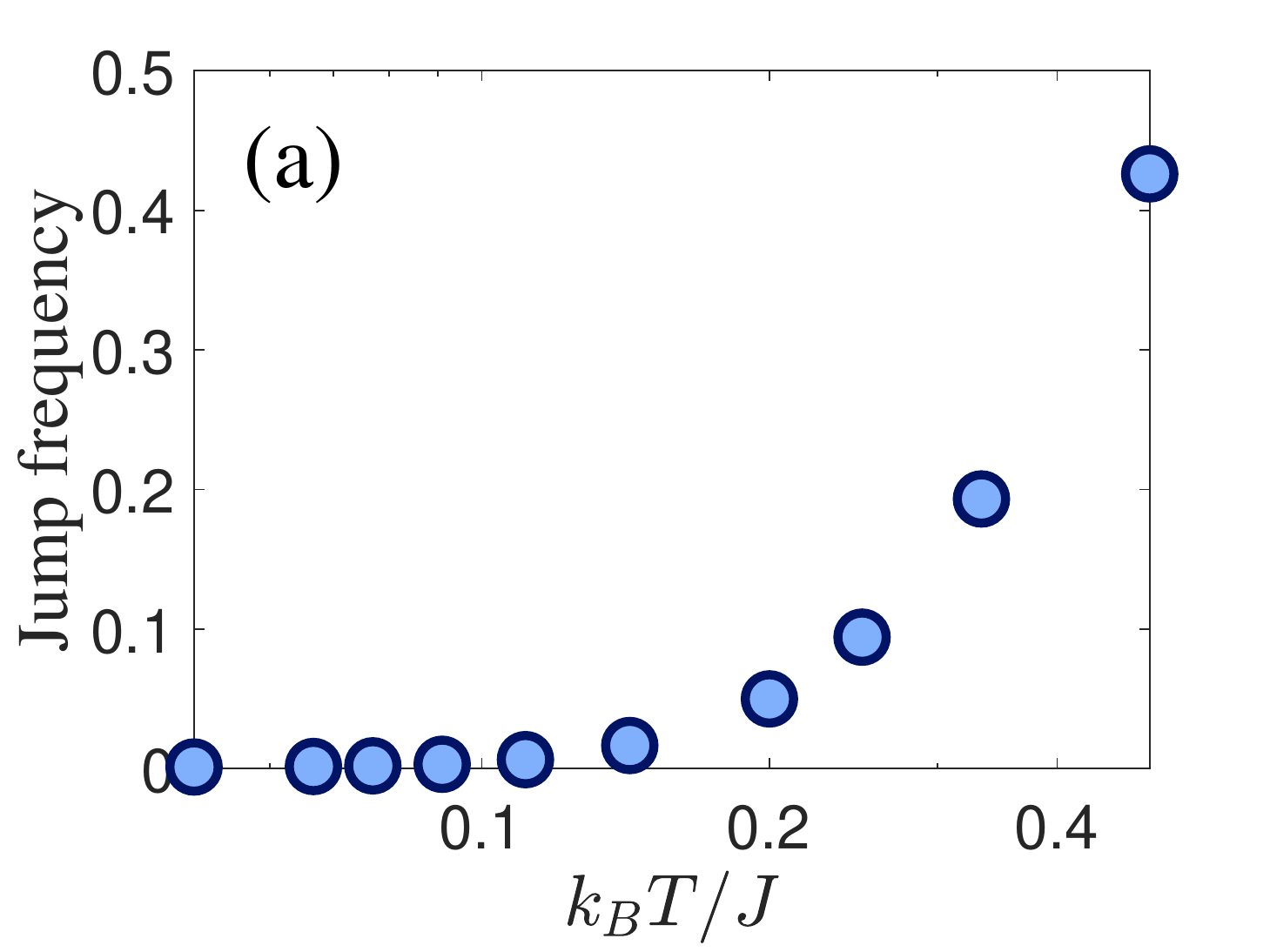}&
\includegraphics[width=.4\linewidth]{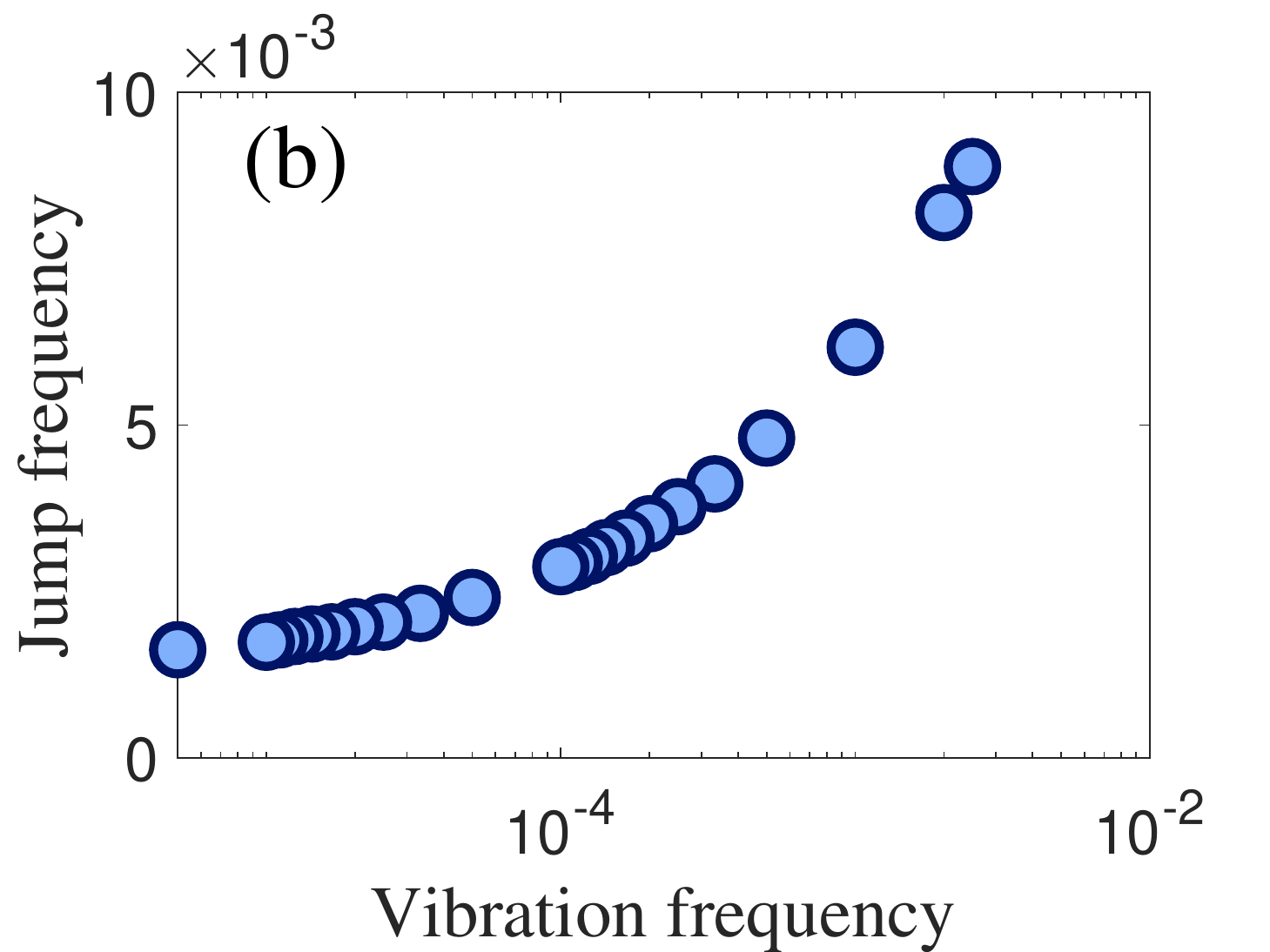}
\end{array}$
\end{center}

\begin{center}$
\begin{array}{ll}
\includegraphics[width=.4\linewidth]{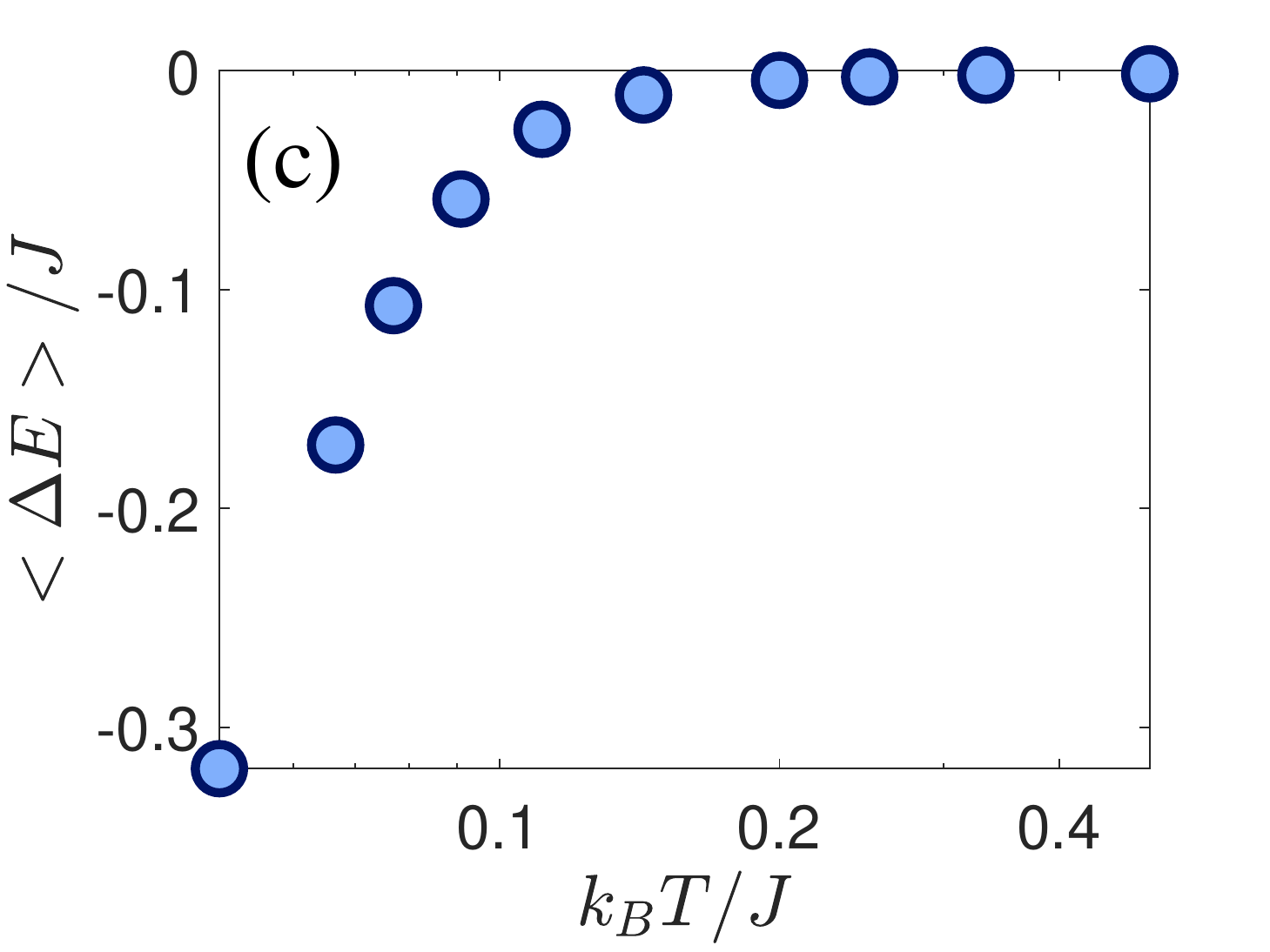}&
\includegraphics[width=.4\linewidth]{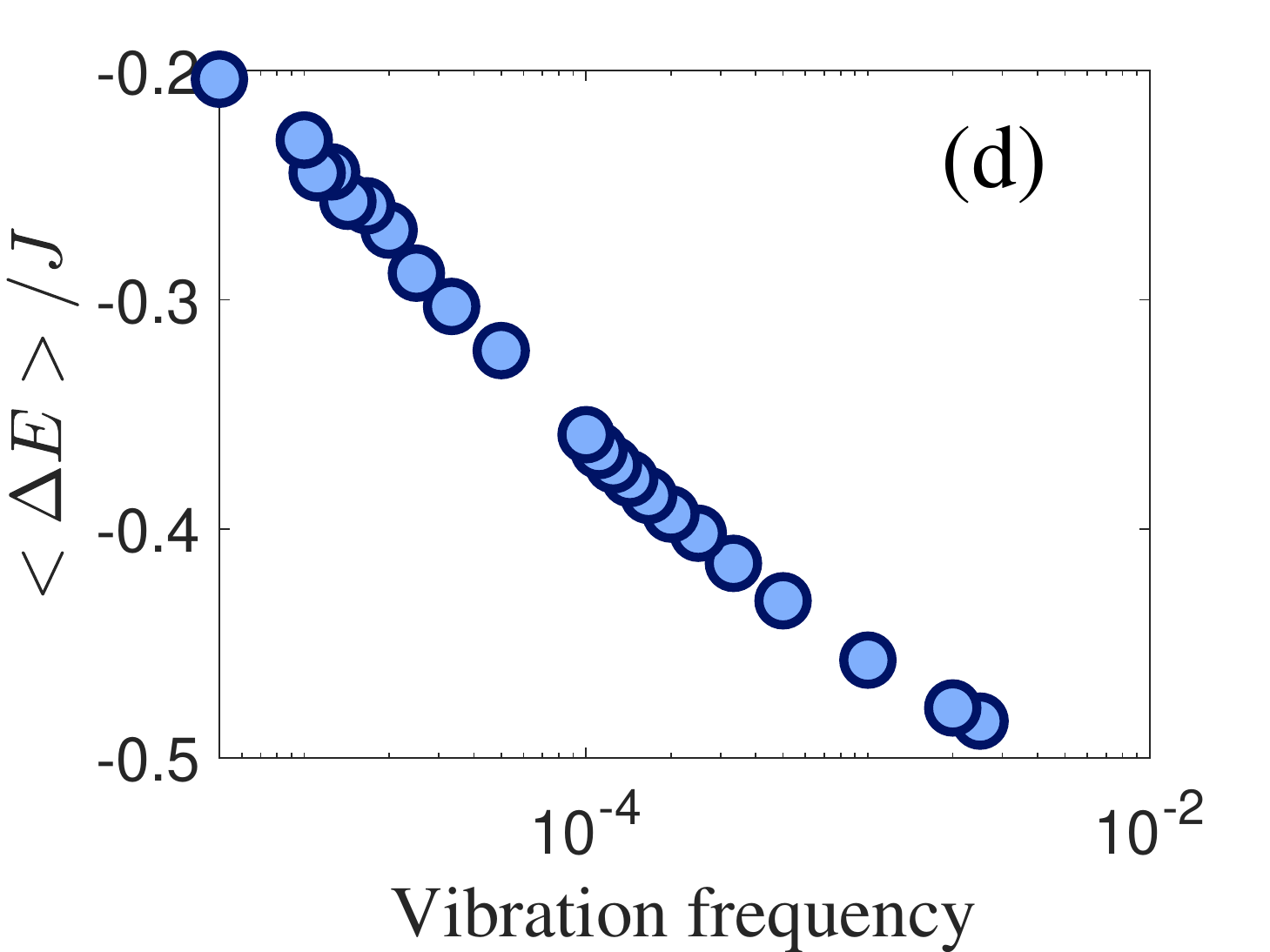}
\end{array}$
\end{center}

\caption{
Single particles behavior during cluster formation. Observed quantities are jump frequency (successful number of transitions divided by total number of KMC steps) and average energetic gain for each successfully transition ($\Delta E =$ energy at the target cell minus the energy at the origin cell). 
\textbf {a} Jump frequency as a function of temperature and no dynamical cluster variations. 
\textbf {b} Jump frequency as a function of vibration frequency and fixed temperature $k_B T/J= 1/15$.
\textbf {c} Energetic gain as a function of temperature and no dynamical cluster variations.
\textbf {d} Energetic gain as a function of vibration frequency and fixed temperature $k_B T/J= 1/15$.
In all the cases the averaging was performed over $100$ realizations. In each realization $260$ particles were present and the measurement time is $10^6$ KMC steps.
}
\label{cluster_dynamics}
\end{figure}

 To better characterize the differences and the similarities of the impact of the two noises, we explore the single-particle behavior. 
 We define jump frequency as the total number of transitions between different Voronoi cells performed by the self-assembling particles, divided by  the measurement, i.e., total number of KMC steps. 
 Panel {\bf{(a)}} of Fig.~\ref{cluster_dynamics} shows that the jump frequency monotonically grows with the temperature when there are no substrate variations. Similar behavior of the jump frequency appears when the temperature is kept fixed, and the vibration frequency is modified, Fig.~\ref{cluster_dynamics} {\bf{(b)}}.
 Yet again, the scales of the panels disclose that the effect of temperature modifications is superior to variations of the vibration frequency. 
 This increase in jump frequency due to the rise of the noise, either a uniform or local, is expected. The jump frequency is associated with a kinetic energy that grows when the temperature is increased. Moreover, the increase of temperature generally allows a system to escape local meta-stable states and reach states with lower potential energy. 
 But when the temperature is too high, the transitions are random, and on average, there is no net energetic gain. Precisely this behavior is observed in panel {\bf{(c)}} of Fig.~\ref{cluster_dynamics}.  
 The average difference between  particle's energy before and after a  transition, i.e., $\langle\Delta E\rangle$, grows with temperature until it saturates at $\langle\Delta E\rangle = 0$.   
 So while the rare transitions at low temperatures occur toward (on average) lower energy states, very frequent transitions do not contribute anything at high temperatures. 
 When average energetic gain, $\langle\Delta E\rangle$, is measured as a function of vibration frequency (for a fixed temperature), a contradiction with  the previously described intuition is observed. Fig.~\ref{cluster_dynamics} {\bf{(d)}} shows that $|\langle\Delta E\rangle|$ always grow with vibration frequency. From this behavior, we can conclude that the effect of local noise due to substrate modifications is different on the single-particle level. 
 Substrate modifications always provide  favorable energetic pathways. These energetic pathways can lead to an improvement in cluster compactness, as is observed in Fig.~\ref{U_vs_temp_or_vibre} {\bf{(b)}}.
 A natural  question arises: If the single-particle transitions are more and more favorable (on average) as the vibration frequency is increased, why is $ U$ a non-monotonic function of the vibration frequency? What mechanism is responsible for the existence of a distinctive minimum of $U$ as a function of vibration frequency? 
 For the case of temperature, the increasing jump frequency and decreasing energetic gain of a single transition balance each other at a specific temperature, thus leading to the observed non-monotonicity in Fig.~\ref{U_vs_temp_or_vibre} {\bf{(a)}}. When the noise is local, the two effects on the single-particle level should contribute to a monotonic growth with the size of the noise, in contradiction with Fig.~\ref{U_vs_temp_or_vibre} {\bf{(b)}}. 
 We will return to this question in the next section. 

\begin{figure}
 \centering
\includegraphics[width=.8\linewidth]{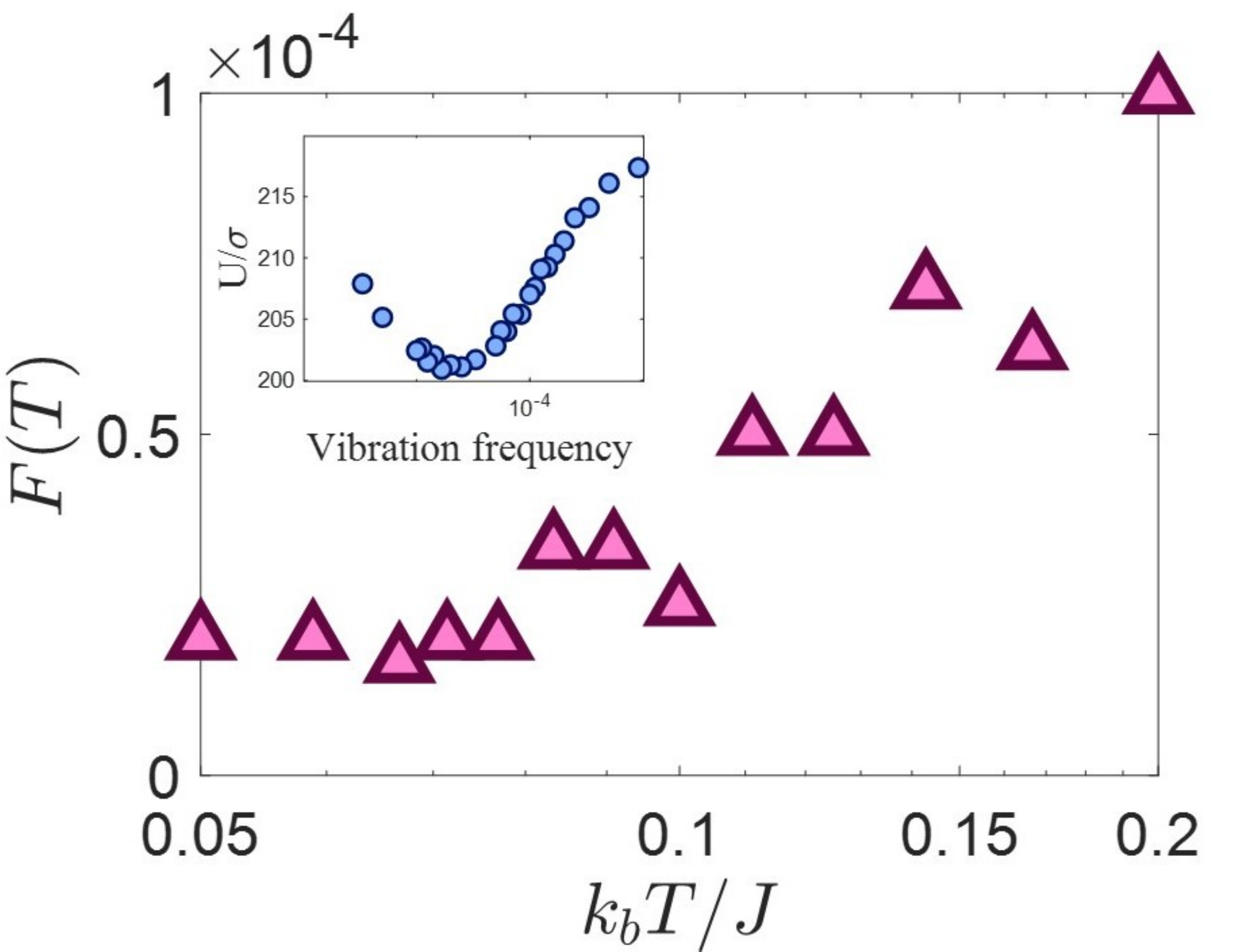}

\caption{Optimal vibration frequency $F(T)$ as a function of temperature. 
For each temperature the minimum of $U$ as a function of vibration frequency was detected (see the inset where $k_B T/J = 0.06$). The vibration frequency that corresponds for this minimum, $F(T)$, is  a growing function of $T$.
For each temperature averaging over $1000$ realizations were performed. Each realization included $300$ particles and the measurement time was $6\times 10^6$ KMC steps.  
}
\label{U_vs_vibration_frequency}
\end{figure}

In Fig.~\ref{U_vs_temp_or_vibre} {\bf{(b)}} we have shown that for $k_B T/J\approx 0.06$, $U$ attains a minimum value when the vibration frequency is  $\approx 3\times 10^{-5}$. 
We define the optimal vibration frequency $ F(T)$ as the vibration frequency for which at temperature $T$ the cluster compactness measure attains a minimum. 
Fig.~\ref{U_vs_vibration_frequency} presents $F(T)$ as a function of $k_BT/J$.
Since both the temperature and the vibration frequency describe the "amount" of noise cast upon the system, we expect some balance between these noises to exist. 
Indeed, for  temperatures below the optimal temperature of Fig.~\ref{U_vs_temp_or_vibre} {\bf{(a)}} additional noise from the vibrations of the substrate can push the system towards the point with optimal temperature. 
Hence, as we approach this optimal temperature, the "amount" of additional noise, i.e., vibration frequency, should decrease.
The behavior in Fig.~\ref{U_vs_vibration_frequency} is quite the opposite. 
It is always beneficial to introduce local noise, and the optimal amount of this noise grows with temperature.
This growth supports the assumption that we raised above regarding different energetic pathways created due to the presence of local noise. 
The impact of the two noises differs quantitatively and qualitatively; they are not interchangeable.

 \section{Discussion}

 This study explored the self-assembly process on top of a 2D amorphous material that lacks periodicity. 
 This lack of periodicity can soften some constraints that limit the self-assembly of crystals by imposing a variation of particle-particle interactions. 
  Previous studies have shown that uniform \cite{Self_Assembly_Shape_Shifting} or local \cite{Bisker_pnas} alternation of particle-particle interaction accelerates self-assembly process. In this line of studies, modification comes from a built-in feature that characterizes each component of the system such as the shape of the component  \cite{Self_Assembly_Shape_Shifting} or the internal state of the component \cite{Bisker_pnas}. In our case, modification of particle-particle interaction appears due to substrate variation, i.e., liquidity.
  Substrate variations modify the local structure of the substrate and, as a consequence, alter the average positions of the particles on top of the substrate, which in turn change the particle-particle interaction. 
  Since these modifications are random, we compare the impact of substrate variations to the effect of temperature, i.e., noise in the system. 
  Similar to thermal fluctuations, substrate modifications stimulate the transition of particles from one location to another.
  But unlike thermal fluctuations, the average energetic gain from such transitions does not wear out when the vibration frequency of the substrate is large. 
  Despite this net energetic gain per particle, the effect on cluster compactness is quite similar for both noises. There is a specific temperature/vibration frequency for which $U$ is minimal, and more noise is destructive. 
  While large thermal fluctuations cause random transitions that destabilize the cluster, large substrate variations can rip the cluster apart by disconnecting local neighbors. 
  These modifications effectively increase particle energy, making the following transitions energetically favorable for the particle but might not be perfect for the cluster formation.
  We can say that while temperature fluctuations facilitate transitions on top of a fixed energy landscape, substrate vibrations cause time-dependent deformations to this landscape. 
  These deformations provide additional energetic pathways toward better self-assembly, up to a specific frequency of deformations.  
  When these deformations of the energetic landscape are present, the system is out of equilibrium. 
  Equilibrium relaxation on the energetic landscape formed during the non-equilibrium period leads to better self-assembly (i.e., more compact cluster). 
  We find that it is always preferable to impose substrate variations, i.e., time-dependent deformations to the energy landscape. When the two noises are applied in a cohort, there is a preferable frequency of variations for any temperature. 
  Moreover, this frequency monotonically grows with temperature. Suggesting that the higher are the thermal fluctuations, the larger amount of substrate variations is needed.

 Thus, behavior on the single-particle level, faster relaxation towards more compact cluster, and consistent improvement of the cluster due to additional frequency of vibrations (that grows with the temperature) advocate for a qualitative difference between the observed behavior and annealing due to temperature.
 Additional heating of the sample is known to produce faster freezing, e.g., the Mpemba effect for which one of the explanations suggests faster search on top of a constant energy landscape occurs due to better initial spread achieved due to pre-heating~\cite{MmpembaArticle}.
 In our case, it is not only that the self-assembly relaxes faster after substrate variations, but it also reaches more compact clusters. 
 The search for the compact cluster occurs not only on a given energy landscape; imposed variations generate a search between different energy landscapes as well. 
 The lack of periodicity of the 2D amorphous substrate allows us to perform this search among different energy landscapes, in the first place.
 When the substrate is not constrained by crystalline order, local modifications are possible. These modifications change the energy landscape and effectively create two processes, one on top of the energy landscape and the other is the transitions of the landscape itself. The observed amplification of self-assembly can appear due to effective cooperation, like in the case of stochastic resonance~\cite{Stochastic_resonance_Luca}.

 When we compare the dynamical range of the effect for thermal fluctuations and substrate variations, we find that thermal fluctuations are superior. 
 Nevertheless,
 the consistency of the effect suggests in favor of promoting and improving the idea of a search among different energy landscapes.  It will be beneficial not only for designers of self-assembly on top of 2D amorphous materials but also for any search process driven by fluctuations on a constant energy landscape, such as Evolution~\cite{EvolutionAmir}  Deep Learning~\cite{DeepLearning}.

{\bf{ACKNOWLEDGMENTS}}
D.S. thanks E. Shimshoni, D.A. Kessler and E. Lazar for fruitful discussions. This work was supported by Israel Science Foundation Grant No. 2796/20.

\bibliography{./deborah_4}

\end{document}